\documentclass[aps,twocolumn,superscriptaddress,shortbibliography]{revtex4-1}
\usepackage{amsmath,amssymb}
\usepackage{graphicx}
\usepackage{color}
\usepackage[dvipsnames]{xcolor}
\usepackage{hyperref}
\usepackage{float}
\usepackage{blindtext}
\usepackage[T1]{fontenc}
\usepackage[utf8]{inputenc}

\usepackage{rotating}
\usepackage{lipsum}
\usepackage{mathtools}
\usepackage{graphicx}
\usepackage{dcolumn}
\usepackage{bm}
\usepackage{amsfonts}
\usepackage{gensymb}
\usepackage{romannum}
\usepackage{changes}
\usepackage{titlesec}

\usepackage{etoolbox} 
\usepackage{lipsum} 
\usepackage[capitalize]{cleveref}

\makeatletter
\appto{\appendix}{%
  \@ifstar{\def\theequation@prefix{A.}}%
          {}%
}
\makeatother

\definechangesauthor[name={HI}, color=orange]{HI}

\newcommand{\bra}[1]{\ensuremath{\left\langle#1\right|}}
\newcommand{\ket}[1]{\ensuremath{\left|#1\right\rangle}}

\newcommand{\mel}[3]{\ensuremath{\left\langle #1 \middle| #2 \middle| #3 \right\rangle}}

\begin{document}
\renewcommand{\thepage}{\arabic{page}}

\title{Realistic tight-binding model for monolayer transition metal dichalcogenides in 1T' structure}

\author{Mengli HU}
\author{Guofu MA}
\author{Chun Yu WAN}
\author{Junwei LIU}
\email{liuj@ust.hk}
\affiliation{Department of Physics, Hong Kong University of Science and Technology, Clear Water Bay, Hong Kong, China}

\date{\today}

\begin{abstract}
Monolayer transition metal dichalcogenides $MX_2$ ($M$ = Mo,W and $X$ = Te, Se, S) in 1T' structure were predicted to be quantum spin Hall insulators based on first-principles calculations, which were quickly confirmed by multiple experimental groups. For a better understanding of their properties, in particular their responses to external fields, we construct a realistic four-band tight-binding (TB) model by combining the symmetry analysis and first-principles calculations. Our TB model respects all the symmetries and can accurately reproduce the band structure in a large energy window from -0.3 eV to 0.8 eV. With the inclusion of spin-orbital coupling (SOC), our TB model can characterize the nontrivial topology and the corresponding edge states. Our TB model can also capture the anisotropic strain effects on the band structure and the strain-induced metal-insulator transition. Moreover, we found that although $MX_2$ share the same crystal structures and have the same crystal symmetries, while the orbital composition of states around the Fermi level are qualitatively different and their lower-energy properties cannot fully described by a single $\bm{ k \cdot p }$ model. Thus, we construct two different types of $\bm{ k \cdot p }$ model for $M$S$_2$,$M$Se$_2$ and $M$Te$_2$, respectively. Benefiting from the high accuracy and simplicity, our TB and $\bm{k \cdot p}$ models can serve as a solid and concrete starting point for future studies of transport, superconductivity, strong correlation effects and twistronics in 1T'-transition metal dichalcogenides.
\end{abstract}


\pacs{}{}
\maketitle

\section{INTRODUCTION}
\
Since proposed in 2005,quantum spin Hall insulators (QSHIs)\cite{kane2005quantum} have attracted tremendous attention due to its unexpected nontrivial topology and potential applications in next-generation information technologies. There are many theoretical predictions of QSHIs in various material systems\cite{hasan2010mz,qi2011topological,bansil2016colloquium}. To date, however, only three material systems have exhibited quantized edge conductance in transport experiments: HgTe quantum wells\cite{bernevig2006ba,konig2007m}, InAs/GaSb quantum wells\cite{liu2008c,knez2011knez} and monolayer WTe$_2$-type transition metal dichalcogenides (TMDCs)\cite{qian2014x,fei2017z,wu2018observation}. Among them, WTe$_2$-type TMDCs are the most promising for potential applications of QSHIs, as they are 2D van der Waals materials.

WTe$_2$-type TMDCs were first predicted to be QSHIs in 2014 \cite{qian2014x} and was experimentally confirmed in WTe$_2$ in 2017 by several independent experiments including transport\cite{fei2017z,wu2018observation}, angle-resolved photoemission spectroscopy (ARPES)\cite{tang2017quantum} and scanning tunneling microscopy (STM)\cite{tang2017quantum,peng2017stm,jia2017direct}. Remarkably, the quantized conductance persists at a high temperature of around 100 K\cite{wu2018observation}. Besides the nontrivial topology, the thin films of WTe$_2$ also possess many other interesting properties including gate-induced superconductivity\cite{Palomaki2018,fatemi2018electrically,hsu2019inversion,xie2020spin} and ferroelectricity\cite{fei2018ferroelectric,sharma2019room}.
Although inversion symmetry is preserved in monolayer 1T'-WTe$_2$, it is broken in the bulk and multilayer thin films, which leads to the existence of Weyl points\cite{sol2015wyle2}, Berry curvature dipole\cite{Xu2018dipole}, and spin-orbit torques\cite{MacNeill2017}. Besides, WTe$_2$ also exhibits a temperature-induced Lifshitz transition\cite{wu2015temperature} and pressure-driven superconductivity\cite{pan2015pressure,kang2015superconductivity}.

In addition to the extensive studies of WTe$_2$, the exploration has also been extended to other related materials. For example, monolayer 1T'-WSe$_2$\cite{chen2018large,pedramrazi2019manipulating,chen2021epitaxial} , 1T'-MoSe$_2$\cite{cheng2019interface}, 1T'-MoTe$_2$\cite{naylor2016monolayer} and 1T'-MoS$_2$\cite{xu2018observation} are successfully fabricated; plus WSe$_2$ and MoS$_2$ show a gapped bulk and conducting edge states in experiments. Observations of topological Fermi arcs\cite{deng2016fermiarc} and superconductivity\cite{Qi2016} in bulk MoTe$_2$ are also reported.

So far, all the experiments about monolayer WTe$_2$ are consistent with each other on the nontrivial topology, and transport experiments clearly show that the bulk states of monolayer WTe$_2$ are insulating\cite{fei2017z,wu2018observation}. However, on the origin of such an insulating behavior, ARPES and STM experiments give contrasting results. ARPES experiments show a non-zero single-particle gap of monolayer WTe$_2$ \cite{tang2017s}, while the STM experiments show that monolayer WTe$_2$ is gapless in the single-particle level, and the gap is from electron-electron correlation pinned at the Fermi level\cite{song2018observation}. Meanwhile, there are some other strange transport behaviors varying with the gating voltage and temperature, which cannot be explained by a single-particle gap\cite{fei2017z,wu2018observation}. A general theoretical picture of all these experimental results is crucial for future studies and applications. Although first-principles calculations can characterize the band structure of 1T'-$MX_2$ \cite{qian2014x}, the huge computational cost makes it incapable to study correlation effects and related properties under various doping/gating, strains and electric fields.

\begin{figure}[htbp]
\includegraphics[width=\columnwidth]{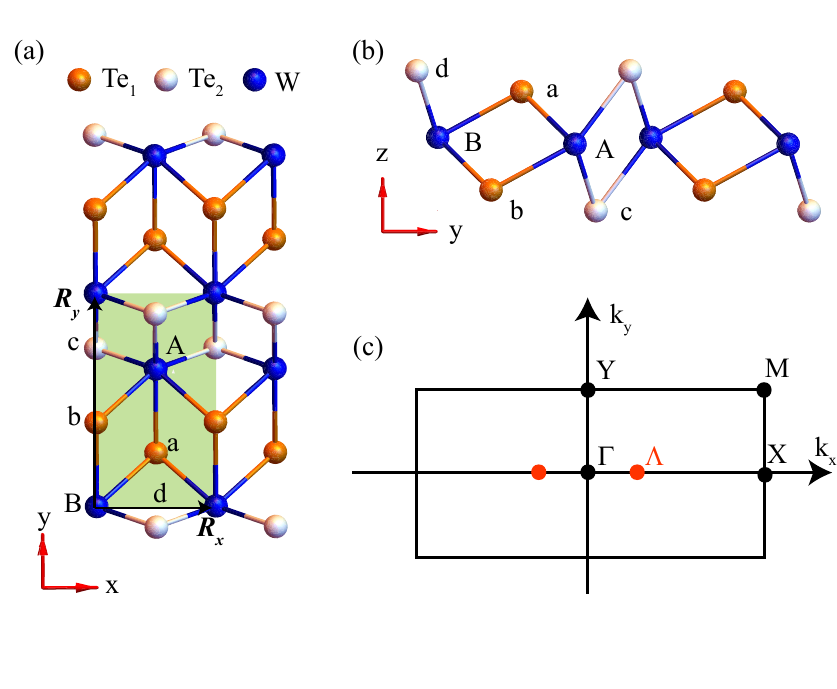}
\caption{\label{Fig.1}Crystal structure of monolayer 1T'-WTe$_2$. (a) top view and (b) side view of 1T'-WTe$_2$. For Te atoms, we divide them into two groups (labeled by Te$_1$ (orange) and Te$_2$ (white)), where two atoms in each group are connected by inversion symmetry. (c) Brillouin zone of monolayer 1T'-WTe$_2$. Other 1T'-$MX_2$ materials own the same lattice structures with different lattice constants.}
\end{figure}

Here, we build a realistic tight-binding model that can capture the low-energy physics of 1T'-$MX_2$ well at the level of first-principles calculations. Different from the previous works\cite{muechler2016topological,ok2018custodial} that can only describe the dispersion along some high symmetry lines, our model can accurately characterize the band structure in the whole Brillouin zone in a large energy window around the Fermi level from -0.3 eV to 0.8 eV for all the monolayer $MX_2$ in 1T' structure. The SOC effects which open the band gaps of Dirac cones are precisely captured in our method.
The SOC drives the system into a QSHI as it is in the Kane-Mele model \cite{kane2005quantum}. Moreover, our TB model can describe the anisotropic effects of strain on the band structure, and can correctly show the strain-induced metal-insulator phase transition. Based on the different orbital components around $\Gamma$ point, we build two types of $\bm{k \cdot p}$ model with fewer parameters. Our results here provide a solid basis for future theoretical studies of transport properties and twistronics in 1T'-$MX_2$.
This paper is organized as follows. In Sec. \ref{spinless TB}, we built two different four-band TB models with different numbers of parameters, and then showed the nontrivial topology and the edge states.
In Sec. \ref{kp model}, we introduced two types of $\bm{ k \cdot p }$ model at $\Gamma$ point.
In Sec. \ref{strain}, the strain effects are included in the TB model. Summary is given in Sec. \ref{summary}.

\section{THE FOUR-BAND TB MODEL}\label{spinless TB}

\begin{table*}[t]
\begin{tabular}{p{2cm} p{2cm} p{2cm} p{2cm} p{2cm} p{2cm} p{2cm} p{2cm} }
\hline\hline
  &$\ \ \ \mu_p$ & $\ \ \ \mu_d$ & $\ \ \ d_1$ & $\ \ \ d_2$ & $\ \ \ d_3$ & $\ \ \ d_4$ &$\ \ \ d_5$\\
  & $\ \ \ p_1$ & $\ \ \ p_2$ & $\ \ \ p_3$ & $\ \ \ p_4$ & $\ \ \ p_{6}$  & $\ \ \ t_1$ & $\ \ \   t_2$  \\
\hline
WTe$_2$  & -1.66528 &\ 0.22665 &\ 0.44406 & -0.18865 &\ 0.05731 & -0.13324 &\ \ \ \ \ -- \\
         &\ 1.03803 &\ 0.34827 &\ 0.17243 & -0.00154 &\ \ \ \ \ -- &\ 0.99303 & -0.03319 \\
MoTe$_2$  & -1.63377 &\ 0.01866&\ 0.28077 & -0.21897 &\ 0.11990 & -0.02540 &\ \ \ \ \ -- \\
          &\ 1.17008 &\ 0.42850 &\ 0.03297 & -0.14288 &\ \ \ \ \ -- &\ 0.88162 & -0.02258 \\
WSe$_2$  & -1.35362 &\ 0.61482 &\  0.27092 &  -0.33119 &\ 0.02467 & -0.03680 &\ 0.02103 \\
		 &\ 0.67497 &\  0.15501 &\ 0.62199 & -0.07087 & -0.27381 &\ 0.50758  & -0.11648 \\
MoSe$_2$ & -1.29397 &\ 0.51532 &\ 0.45815 & -0.20800 &\ 0.09843  &  -0.05272 &\ 0.09275 \\
		 &\ 0.63585 &\ 0.15681 &\ 0.10656 & -0.04924 & -0.00158 &\ 0.80408  & -0.06020 \\
WS$_2$  & -1.35183 &\ 0.63848 &\  0.23062 & -0.32906 &\ 0.03482 & -0.04138  &\ 0.01661 \\
	    &\ 0.64428 &\ 0.12264 &\ 0.65028  & -0.07014 & -0.28547 &\ 0.49754  & -0.04101  \\
MoS$_2$  & -1.36700 &\ 0.62198  &\ 0.20933 & -0.32600 &\ 0.03672 & -0.02389 &\ 0.02699\\
         &\ 0.63741 &\ 0.17775 &\ 0.63298 & -0.08625 & -0.27196 &\ 0.53014 & -0.08643 \\
\hline\hline
\end{tabular}
\caption{Fitted $\uppercase\expandafter{\romannumeral1}$-TB model parameters for all six $MX_2$ in units of eV.}
\label{TABLE 1}
\end{table*}

\begin{table*}[t]
\begin{tabular}{p{1.5cm} p{1.5cm} p{1.5cm} p{1.5cm} p{1.5cm} p{1.5cm} p{1.5cm} p{1.5cm} p{1.5cm} p{1.5cm} p{1.5cm}}
  \hline\hline
   &\ \ \ $\mu_p$ & \ \ \ $\mu_d$ & \ \ \ $d_1$ & \ \ \ $d_2$     & \ \ \ $d_3$  & \ \ \ $d_4$  & \ \ \ $d_5$  & \ \ \ $d_6$  & \ \ \ $p_1$ &\ \ \ $p_2$   \\
   & \ \ \ $p_3$   & \ \ \ $p_4$ & \ \ \ $p_5$ & \ \ \ $p_{6}$  & \ \ \ $t_1$ & \ \ \ $t_2$ & \ \ \ $t_3$ & \ \ \ $t_4$ & \ \ \ $t_{5}$ & \ \ \ $t_{6}$\\
  \hline
  WTe$_2$ & -1.89759 &\ 0.26199 &\ 0.40510 & -0.22072 & -0.07897 &\ 0.17756 &\ 0.02538 &\ \ \ \ \ --  & 1.13760 &\ 0.31121 \\
          &\ 0.76753 & -0.02325 &\  \ \ \ \ --  & -0.31933 &\ 0.95946 & -0.25719 &\ 0.37810 & -0.15573 &\  \ \ \ \ -- &\  \ \ \ \ --  \\
  MoTe$_2$ & -2.03758 &\ 0.03843 &\ 0.31260 & -0.23191 & -0.09447 &\ 0.17815 &\ 0.02140 &\ \ \ \ \ --  & 1.23257 &\ 0.31056 \\
           &\ 0.69232 & -0.08343 &\  \ \ \ \ --  & -0.29010 &\ 0.81744 & -0.23141 &\ 0.39989 & -0.09774 &\  \ \ \ \ -- &\ \ \ \ \ --  \\

  WSe$_2$ & -1.3263 & \  0.68423 &\  0.24922 & -0.26472&\  0.06792 & -0.00247 &\  0.02786& -0.00782&\  0.68267 & \ 0.22103 \\
          &\  0.66442&\  0.00497& -0.05935 & -0.26452 &\  0.75205 & -0.08788 &\ 0.02258& -0.14976  & -0.07251  & -0.02152 \\
  MoSe$_2$ & -1.94659&\  0.27827 &\  0.33844& -0.28305& -0.1167 &\  0.20624&\  0.02506&\  0.02712&\  1.03571 &\ 0.25439 \\
           &\  0.78674&  -0.0228 &  -0.02966& -0.33199 &\  0.89168 & -0.29952&\ 0.37373&  -0.11279&\  0.00868& -0.00936  \\
  WS$_2$    & -1.34342 &\ 0.87386 &\ 0.27347 & -0.35576 &\  0.05040 & -0.04117 &\ 0.05692 & -0.01395 &\ 0.63408 &\ 0.11487 \\
  			&\ 0.66955 &\ 0.01900 & -0.09228 & -0.29918  &\ 0.64972 &\ 0.00212  &\ 0.09219 & -0.03392 & -0.07035 &\ 0.03078 \\
  MoS$_2$   & -1.33942  &\ 0.65475  &\ 0.22663  & -0.30787  &\  0.05831 &\ 0.04386 &\ 0.00695   & -0.00742 &\ 0.64578 &\ 0.17477 \\
  			&\ 0.66761   & -0.01281  & -0.05326  & -0.28997  &\ 0.70452  & -0.05753  &\ \ \ \ \ --  & -0.05290   & -0.06017 &\ 0.09061 \\
  \hline\hline
\end{tabular}
\caption{ Fitted $\uppercase\expandafter{\romannumeral2}$-TB model parameters for all six $MX_2$ in units of eV.}
\label{TABLE 2}
\end{table*}

We first analyze the crystal symmetries of monolayer $MX_2$ in 1T' structure and then build a four-band TB model based on orbital contributions from first-principles calculations. The parameters in the TB models are further fitted according to the band structure in certain energy ranges around the Fermi level.

\begin{figure}[htbp]
\includegraphics[width=\columnwidth]{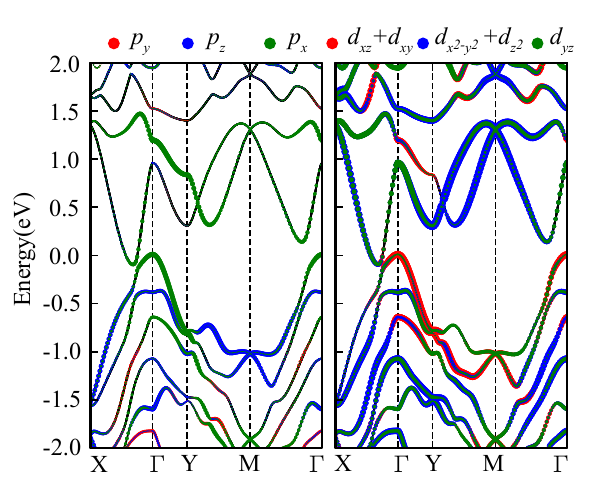}
\caption{\label{Fig.2} Orbital projections for monolayer 1T'-WTe$_2$ from first-principles calculations. Marker size represents the weight of a specific orbital. Left panel is projection results from $p$ orbitals of Te$_1$. Red for $p_y$, blue for $p_z$ and green for $p_x$. Right panel is the result of contributions from $d$ orbitals of W atoms, red for $d_{xz} + d_{xy}$, blue for $d_{x^2-y^2} + d_{z^2}$ and green for $d_{yz}$.}
\end{figure}

As shown in Fig. \ref{Fig.1}(a), monolayer 1T'-$MX_2$ has a rectangular unit cell with two $M$ atoms and four $X$ atoms. The lattice vectors are labeled by $\bm{R_x}$ and $\bm{R_y}$ with $z$ along the perpendicular direction, and all the unit cells can be uniquely labeled by the vector $\bm{R}_{mn}$ = $m\bm{R_x}$ + $n\bm{R_y}$, with $m,n$ $\in$ $\mathbb{Z}$.The monolayer 1T'-$MX_2$ belongs to a nonsymmorphic space group, containing the following symmetries: 1. Translation $\bm{T_R}$; 2. Glide mirror $\bar{M}_x$: $(x,y,z)\rightarrow (-x,y,z)+\bm{R_x}/2$; 3. Screw rotation $\bar{C}_{2x}$: $(x,y,z)\rightarrow (x,-y,-z)+\bm{R_x}/2$.
There is also an inversion symmetry $P$: $(x,y,z)\rightarrow(-x,-y,-z)$, which can be written as the product of glide mirror and screw rotation $P=\bar{M}_x\bar{C}_{2x}$. For clarity, four $X$ atoms are divided into two types, labelled as $X_1$ (atoms a and b) and $X_2$ (atoms c and d), where two atoms in each type are connected by inversion symmetry.

The nontrivial topology of monolayer 1T'-$MX_2$ comes from the Peierls distortion and the strong SOC\cite{qian2014x}. The Peierls distortion along the $x$ axis induces a large band inversion at $\Gamma$ point, forming two gapless Dirac cones which are protected by the crystal symmetries (the $\Lambda$ points in Fig. \ref{Fig.1}(c)) along $\Gamma-X$. Moreover, the SOC can open band gaps at the Dirac points and drives monolayer 1T'-$MX_2$ into a Z$_2$ topological insulator, which is exactly the same as the Kane-Mele model for graphene\cite{kane2005quantum}. Here, we first build the TB model without SOC.

\subsection{TB model without SOC}

\begin{figure*}[htbp]
\includegraphics{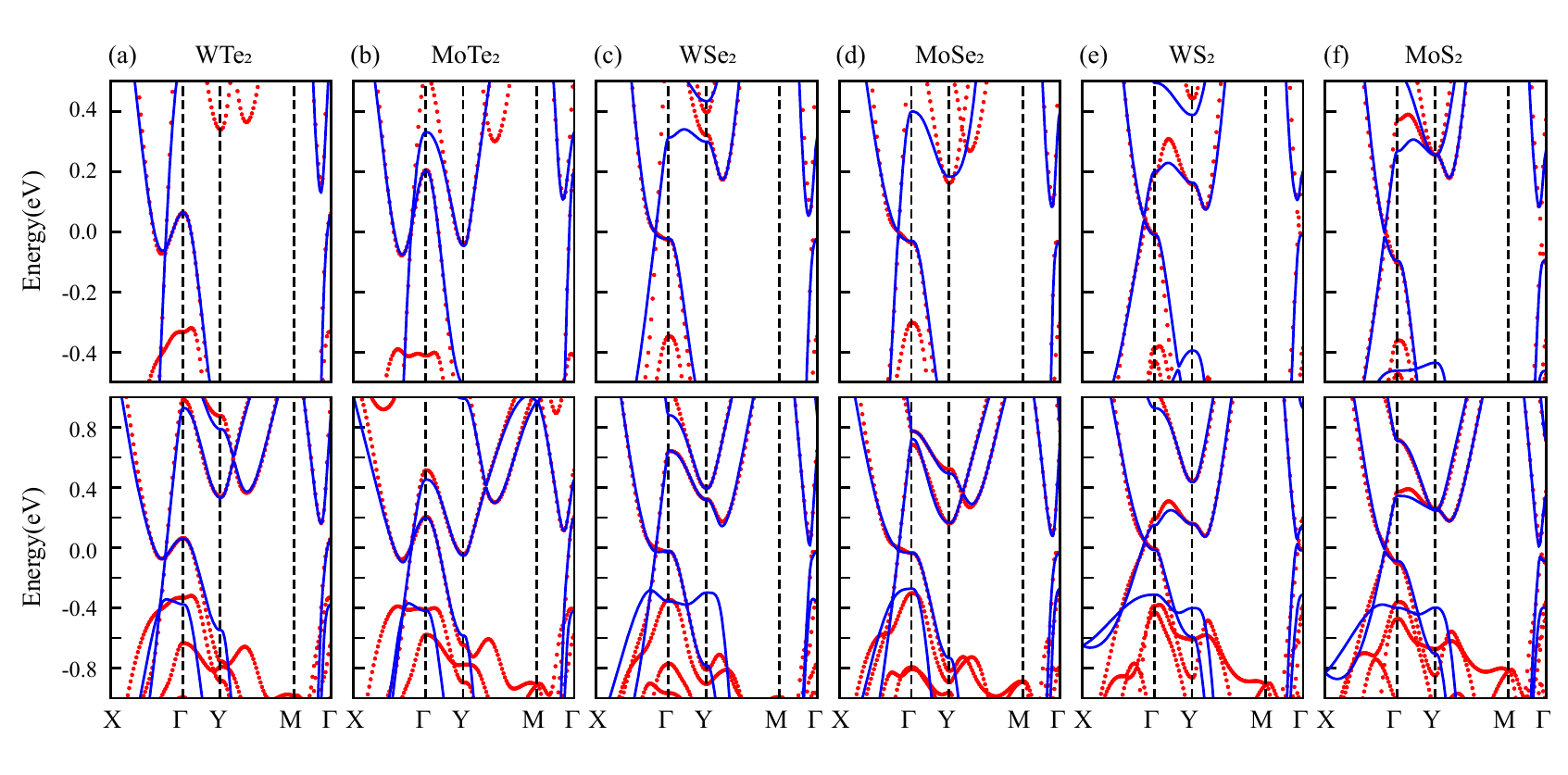}
\caption{\label{Fig.3} The $\uppercase\expandafter{\romannumeral1}$ - and $\uppercase\expandafter{\romannumeral2}$-TB fitted band structures of $MX_2$ with the corresponding first-principles calculations. Red dots are the first-principles calculation results and black solid lines are the fitted TB results. (a-f) Upper panel are the $\uppercase\expandafter{\romannumeral1}$-TB results with energy region from -0.5 eV to 0.5 eV; (a-f) Lower panel are the $\uppercase\expandafter{\romannumeral2}$-TB results with energy region from -1.0 eV to 1.0 eV.}
\end{figure*}

The band structures of monolayer 1T'-$MX_2$ can be accurately calculated by first-principles calculations\cite{qian2014x,inversion,tang2017quantum}.
Our first-principles calculations are performed in the framework of density functional theory as implemented in the VASP package\cite{vasp} with projector augmented wave (PAW) method\cite{paw}. The exchange-correlation functional is Berdew-Burke-Ernzerhof (PBE)\cite{PBE}. The Brillouin zone is sampled in a $7 \times 14 \times 1$ Monkhorst-Pack k-point grid and a large vacuum region of more than 15 \AA\ is used in the $z$ direction to minimize image interactions from the periodic boundary condition. The energy cutoff of the plane wave basis is 400 eV. The relaxed lattice constants are consistent with the previous works\cite{qian2014x,inversion,tang2017quantum}.

To be more specific, we focus on WTe$_2$ in the following part. As shown in Fig. \ref{Fig.2}, the states in the energy window from -0.5 eV to 1.0 eV are mostly contributed by $\{d_{x^2-y^2}, d_{z^2}\}$ orbitals of W atom and $p_x$ orbital of Te$_1$ atom, which belong to the irreducible representations (IRs) $A_g\{d_{x^2-y^2},d_{z^2},d_{yz}\}$ and $A_u\{p_{x}\}$ of the group $C_{2h}$, respectively.
Thus we choose two $d$ orbitals from W$_A$ and W$_B$, and two $p$ orbitals from Te$_{1a}$ and Te$_{1b}$ as our bases, which are labeled as $\ket{\phi_{l}}$ ($l = A,B,a,b$). The TB Hamiltonian without SOC in the bases $\phi_{l}$ is $H = \sum_{ll',mm',nn'}t_{ll',mm',nn'}\ket{\phi_{lmn}}\bra{\phi_{l'm'n'}}$. The hopping parameter $t_{ll',mm',nn'}$  between two orbitals at $\bm{R_{m,n} + \tau_l}$ and $\bm{R_{m',n'} + \tau_{l'}}$ can be calculated as $t_{ll',mm',nn'}=\mel{\phi _ {l}(\bm{r - R_{mn} - \tau_l})} {\hat H} {\phi _{l'}(\bm{r - R_{m'n'} - \tau_{l'}}) }$, where $\tau_{l}$ is the relative position of an atom in the unit cell.

By including enough hopping parameters, we can get an exact TB model, but it will make the TB model extremely complicated and difficult to use. To strike a balance between simplicity and accuracy, we introduce two TB models involving different numbers of hoppings. After considering the symmetry constraints on different hopping parameters (more details in Appendix A), all the matrix elements in the four-band TB Hamiltonian are
\begin{equation}
\begin{aligned}
H_{11/22}&= \mu_d+2(d_2+2d_5\cos k_y)\cos k_x+2d_4\cos k_y, \\
H_{33/44}&= \mu_p+2(p_1+2p_{6}\cos k_y)\cos k_x+2p_3\cos k_y, \\
H_{12}&=(1+e^{i k_x})(d_3+d_1e^{i k_y}+d_6e^{-i k_y}),   \\
H_{13}&=-2i(\sin k_x)(t_2+t_4{e^{ik_y}}+t_{5}{e^{-ik_y}}),    \\
H_{14}&=(1 - e^{ i k_x})(-t_1 + t_3e^{i k_y} + t_{6}e^{-i k_y}),   \\
H_{23}&=(1 - e^{-i k_x})(t_1 - t_3e^{-i k_y} - t_{6}e^{i k_y}),  \\
H_{24}&=-2i(\sin k_x)(t_2 + t_4{e^{-ik_y}} + t_{5}{e^{i k_y}}),   \\
H_{34}&=(1+e^{i k_x})(p_2 + p_4 e^{-i k_y} + p_5e^{i k_y}),
\end{aligned}
\end{equation}
where $\mu_{p,d}$ is the on-site energy and $\{p,d,t\}_j$ represents the $j$-th nearest hopping parameters within the $p$ or $d$ orbital or between the $p$ and $d$ orbital, respectively.

In experiments\cite{song2018observation}, the 1T'-WTe$_2$ is usually heavily electron doped, thus we fit the TB model with a focus on the conduction bands with a energy region from -0.3 eV to 0.8 eV. The fitted parameters for all the six $MX_2$ are shown in the Table \ref{TABLE 1} and \ref{TABLE 2}, and the fitted results are shown in Fig. \ref{Fig.3}. Interestingly, for WTe$_2$, and MoTe$_2$, the parameters $d_5$ and $p_6$ turn out to be very small and can be safely ignored, which might be due to the semimetallic property of $M$Te$_2$.

\subsection{SOC effects} \label{SOC TB}

In a fermion system, SOC is necessary to get the nontrivial topology. To describe the effects of SOC, we first extend the spinless bases to the spinful bases $\ket{\phi_{l,\alpha}} (l = A,B,a,b; \alpha = \uparrow, \downarrow)$, and the effects of SOC can be taken into account as $H^{soc}_{\alpha\alpha'} = i\sum_{mn} (\bm{E}_{mn}\times \bm{r}_{mn})\cdot \bm{\sigma}_{\alpha\alpha'} \ket{\phi_{lm,\alpha}} \bra{\phi_{l'n,\alpha'}}$, where $\bm{r}_{mn}$ is the vector connecting two atoms at $\bm{r}_m$, and $\bm{r}_n$, $\bm{E}_{mn}$ is the electric field between them, and $\bm{\sigma}$ are the Pauli matrices for spin.

The Hamiltonian with SOC is time-reversal and inversion invariant. Therefore, the band structure is two-fold degenerate over the whole BZ. The main effect of SOC is the opening of the band gap of the Dirac cones at $\Lambda$ points as depicted in Fig. \ref{Fig.1}, which can be captured in our TB model by only considering the nearest off-site SOC. Considering all the symmetry constraints, the full TB model with SOC is
\begin{widetext}
\begin{equation}
\begin{aligned}
H(\bm{k}) = H^{0}(\bm{k}) + \left( g_y \sigma_y + g_z \sigma_z \right) \left[ - l_{1} s_y \tau_z + l_{2}s_x((1+\cos k_x) \tau_y + \sin k_x \tau_x )\right],
\end{aligned}
\end{equation}
where $H^0$ represents the spinless TB Hamiltonian obtained above. $\sigma_i$, $s_i$, and $\tau_i$ stand for the spin, orbital, and sublattice degree of freedom respectively. $l_{1}$ and $l_{2}$ are the SOC strength parameters. It is noted that the $\sigma_x$ related terms are allowed if we consider long-range off-site SOC.
\end{widetext}

\begin{table}
\begin{tabular}{p{1cm} p{1.5cm} p{1.5cm} p{1.5cm} p{1.5cm}}
\hline\hline
      &$l^{\uppercase\expandafter{\romannumeral1}}_{1}$&$l^{\uppercase\expandafter{\romannumeral1}}_{2}$&$l^{\uppercase\expandafter{\romannumeral2}}_{1}$  &$l^{\uppercase\expandafter{\romannumeral2}}_{2}$ \\
\hline
 WTe$_2$  & -0.00666 & -0.04892 & -0.02943 & -0.06564\\
 MoTe$_2$ & \ 0.00699 &\ 0.03192&\ 0.01129&\ 0.04041\\
 WSe$_2$  & \ 0.03193&\ 0.01195& -0.00309&\ 0.05150\\
 MoSe$_2$ & \ 0.00118&\ 0.02147&\ 0.00505& -0.02200\\
 WS$_2$   & \ 0.03000 &\ 0.03000  &\ 0.02000 &\ 0.05000\\
 MoS$_2$  &\ 0.00325& -0.01647&\  0.00861& -0.01021 \\
\hline\hline
\end{tabular}
\caption{Fitted SOC parameters of the $\uppercase\expandafter{\romannumeral1}$- and $\uppercase\expandafter{\romannumeral2$}-TB models (in units of eV) for all six $MX_2$.}
\label{TABLE 3}
\end{table}

\begin{figure*}
\includegraphics{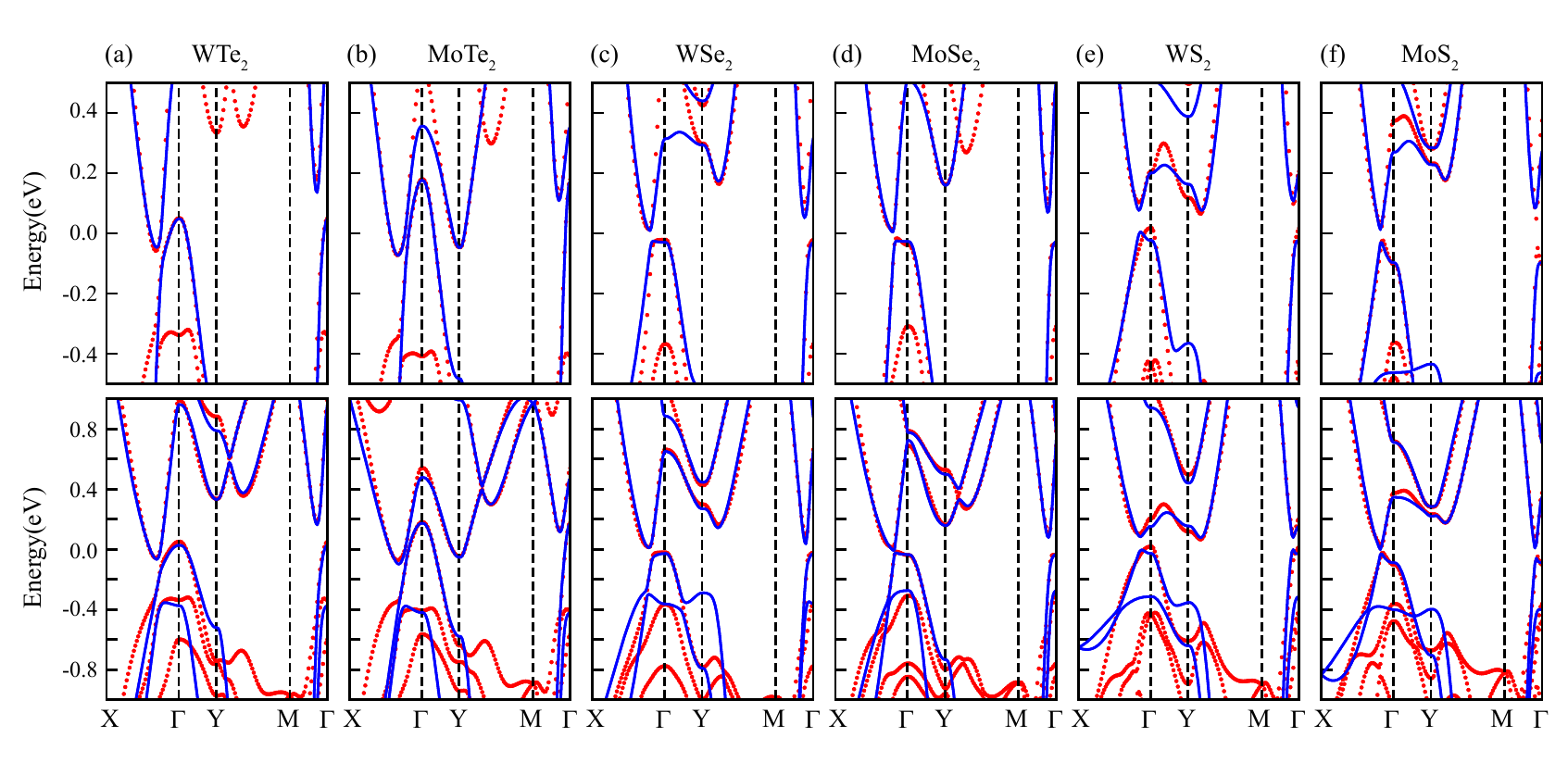}
\caption{\label{Fig.4} The $\uppercase\expandafter{\romannumeral1}$- and $\uppercase\expandafter{\romannumeral2}$-TB fitted band structures of $MX_2$ with SOC. Red dots are first-principles results and black solid lines are fitted TB results. (a-f) Upper panel are the $\uppercase\expandafter{\romannumeral1}$-TB results with SOC; (a-f) Lower panel are the $\uppercase\expandafter{\romannumeral2}$-TB results with SOC.}
\end{figure*}

Without loss of generality, we consider the $\sigma_y$ and $\sigma_z$ terms with the same strength ($g_y=g_z=1$). In Table \ref{TABLE 3}, two sets of independent SOC parameters are listed for the two TB models. The parameters are fitted based on the previous spinless TB model, and the fitted results are shown in Fig. \ref{Fig.4}. Clearly, the SOC opens a gap for the Dirac cone along $\Gamma - X$ around the Fermi level for all $MX_2$ and it drives $M$Se$_2$ and $M$S$_2$ to be an insulators with full fundamental gaps, while it drives $M$Te$_2$ to be a semimetal with a negative fundamental gap. It is worth noting that the $Z_2$ topological invariant is still well-defined for all the $MX_2$ since the direct gap is non-zero due to SOC.

\subsection{Non-trivial topology in $MX_2$} \label{symmetry}

The little group of different momenta are always distinct from the space group of the given material. For 1T'-$MX_2$, four time-reversal-invariant momenta $\Gamma,\ X, \ Y,$ and $M$ belong to $C_{2h}$, the momenta along two high symmetry lines $X - \Gamma$ and $Y - M$ belong to $C_{2}$, the momenta along $\Gamma - Y$ belong to $C_{s}$, and all the other momenta including those along $\Gamma - M$ belong to a trivial symmetry group with only an identity operation.
\begin{figure}
\includegraphics[width=\columnwidth]{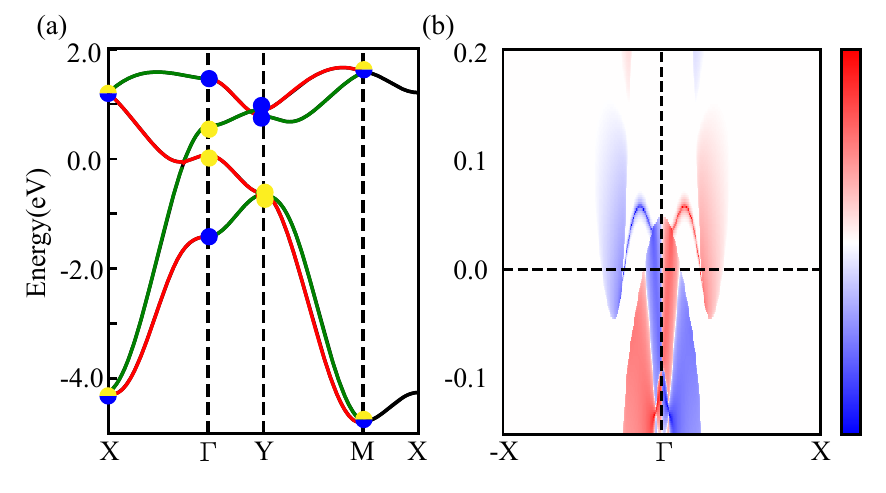}
\caption{\label{edge} (a) The band structure of monolayer 1T'-WTe$_2$ along high symmetry lines from the spinless fitted $\uppercase\expandafter{\romannumeral1}$-TB model. Here we denote parities of states at time-reversal invariant momenta with blue and yellow dots and different characters of states along high symmetry lines by red and green lines. (b) Edge spin-polarized density of states along the $\bm{R_x}$ direction.}
\end{figure}

Different from the simple point group symmetry operation, $\bar{M}_x$ and $\bar{C}_{2x}$ are associated with a translation operation, thus their detailed matrix representations will depend on the momentum. In the bases of \{$\ket{\phi_{A,\bm{k}}},\ket{\phi_{B,\bm{k}}},\ket{\phi_{a,\bm{k}}},\ket{\phi_{b,\bm{k}}}$\}, they are written as
\begin{equation}
\begin{aligned}
  \bar{C}_{2x}(k_x) &=  s_0 \left( \begin{matrix}
    0 & e^{-ik_x} \\ 1 & 0 \end{matrix}
  \right ) , \\
   \bar{M}_{x}(k_x) &=  s_z \left( \begin{matrix}
    1 & 0 \\ 0 & e^{-ik_x} \end{matrix}
  \right ) , \\
   P &= \bar{C}_{2x}(0)\bar{M}_{x}(0).
\end{aligned}
\end{equation}

One of the most profound consequences of these non-symmorphic symmetries is that the degeneracy of the states at different momenta might be different even if their momenta belong to the same little group. For example, both $\Gamma$ and $X$ belong to $C_{2h}$, while $\bar{M}_x$ can guarantee a double degeneracy at $X$ but not at $\Gamma$. It is because in the sublattice degree of freedom, $\bar{M}_x$ is $\tau_z$ at $X$, which anti-commutes with $P=\tau_x$, while $\bar{M}_x$ is $\tau_0$ at $\Gamma$, which commutes with $P=\tau_x$. Thus, under these symmetry constraints, only $\tau_0$ is allowed in Hamiltonian at $X$ leading to the double degeneracy in the sublattice degree of freedom. A similar argument can be applied for the states along $X-M$ where the double degeneracy is protected by $PT$ and $\bar{M}_x$ as shown in Fig. \ref{edge}(a). The two bands around Fermi level along $\Gamma-X$ line belong to different irreducible representations, thus there is a band crossing forming a Dirac cone. With SOC, these two bands will belong to the same representation of the double group, and the band crossing becomes an anti-crossing, leading to a gap opening for the Dirac cone and drives $MX_2$ into a 2D topological insulator.

The nontrivial topology can be verified by directly calculating the $Z_2$ index using the Fu-Kane formula $(-1)^{\upsilon} = \Pi_{i} \Pi_{m = 1}^N \xi_{2m}(\Gamma_i)$, where $N$ is the number of occupied states and $\xi_{2m}(\Gamma_i)$ is the parity of states at different time-reversal-invariant momentum $\Gamma_i$ \cite{inversion}. In Fig. \ref{edge}(a), we show the parities for states at four time-reversal-invariant momenta of WTe$_2$ calculated by our TB model. The $Z_2$ index can be easily checked to be 1.

Besides the nonzero $Z_2$ topological invariant, we also checked the edge states based on our TB models. As shown in Fig. \ref{edge}(b), there are two spin-polarized edge states across the band gap connecting the conduction and the valence bands. For a pair of states at different momenta related by the time-reversal symmetry, the spin polarization must be opposite in all directions. For a pair of states at different momenta related by the mirror symmetry $M_x$, the $x$-component of spin polarization must be the same. Thus, for the edge states, the spin must be in the $y-z$ plane.(Detail
derivation in Appendix C) Our TB models indeed describe these properties correctly as shown in the Fig. \ref{edge}(b)($z$ component) and Appendix B ($y$ component).

It is worth noting that the underlying mechanism of the nontrivial topology here is exactly the same as the Kane-Mele model for graphene with SOC, where the band inversion happens without SOC and results in two Dirac cones protected by crystalline symmetries, and the SOC opens the band gaps for the Dirac cones to drive the Dirac semimetal into a 2D topological insulator.

\section{ \textit{k} $\cdot$ \textit{p} model} \label{kp model}

Many properties of a fermion system like transport are determined by the states around the Fermi surface. For a semiconductor, the states around the Fermi surface are usually centered around some high-symmetry momenta, and their properties can be described by a simpler $\bm{ k \cdot p }$ perturbation model. All the allowed terms in a $\bm{k\cdot p}$ model can be analyzed by considering all the symmetries in the little group of the high-symmetry momentum. Besides, for states belonging to different irreducible representations, the $\bm{k\cdot p}$ models will be different.
\begin{figure}
\includegraphics[width=\columnwidth]{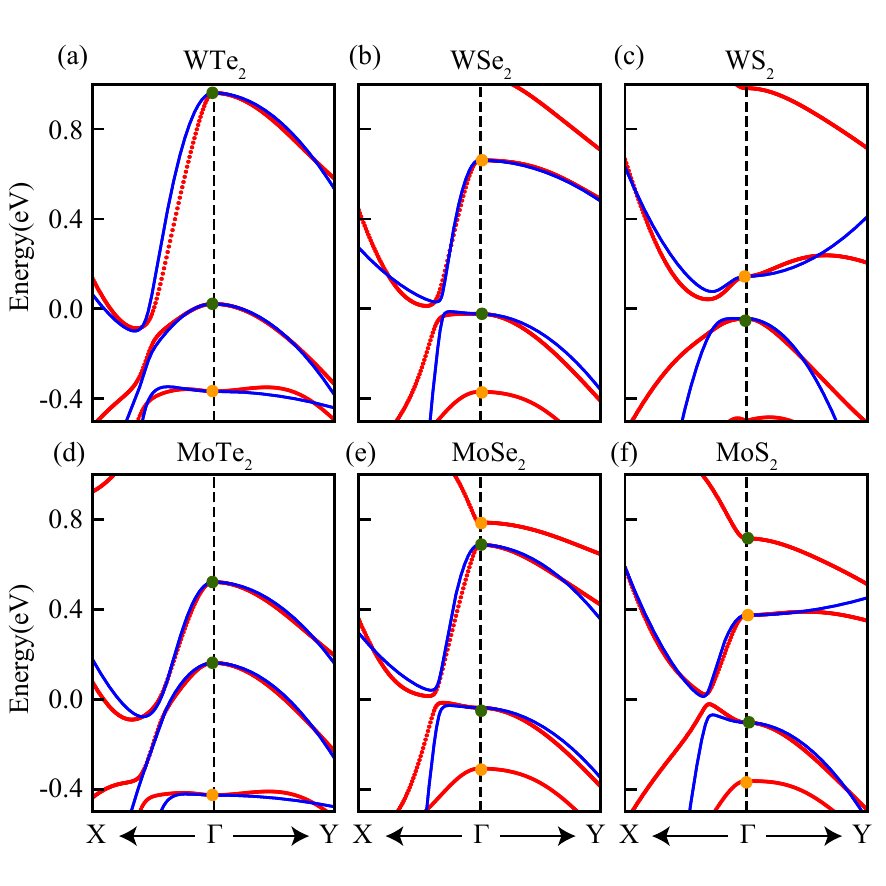}
\caption{\label{Fig.kp} Fitted $\bm{k \cdot p}$ model results for $MX_2$. (a-f) Red dots are from first-principles calculations and blue solid lines are the fitted $\bm{k \cdot p}$ model results.
Green and orange solid dots at $\Gamma$ point represents the even and odd parity, respectively. The fitting region is around $\Gamma$ point and the band inversion momenta.}
\end{figure}
The space group are the same for all six $MX_2$ materials, yet the orbital components around the fermi level are not all identical. At $\Gamma$ point, all six materials present even parity at the valence band maximum (VBM) and different parities at the conduction band minimum (CBM). For $M$Te$_2$, the CBM parity is even and far away(>0.1 eV) from the next odd parity conduction band. It causes failure to capture symmetry and topological properties if only two same parity bases are considered. Therefore, we also consider the state below VBM and build a three-orbital $\bm{k\cdot p}$ model for $M$Te$_2$.
Specifically, we use \{$\ket{\phi_{p_x},\uparrow},\ket{\phi_{p_x},\downarrow},\ket{\phi_{d_{xz},\uparrow}},\ket{\phi_{d_{xz}},\downarrow}$\} as the bases for $M$Se$_2$ and $M$S$_2$, and use \{$\ket{\phi_{d_{z^2}},\uparrow},\ket{\phi_{d_{z^2}},\downarrow},\ket{\phi_{d_{xz}},\uparrow},\ket{\phi_{d_{xz}},\downarrow},\ket{\phi_{p_{y},\uparrow}},\ket{\phi_{p_{y}},\downarrow}$\} for $M$Te$_2$ to construct the $\bm{ k \cdot p }$ model, respectively.

For $M$Se$_2$ and $M$S$_2$, the symmetry operations at $\Gamma$ point are $M_x = i s_z \sigma_x$, $P = s_z \sigma_0$ and time-reversal symmetry operation which reads $T = i s_0 \sigma_y K$. For $M$Te$_2$, symmetry operations at $\Gamma$ point are $P = (\lambda_0 + 2\sqrt{3}\lambda_8) \sigma_0$, $M_x = i (\lambda_0 + 3 \lambda_3 -\sqrt{3}\lambda_8) \sigma_x$ and $T = i \lambda_0 \sigma_y K$. $s_i$ and $\lambda_i$ are Pauli and Gell-Mann matrices for orbital degree of freedom; $\sigma_i$ are Pauli matrices for spin degree of freedom and $I_{4/6}$ is an identity matrix. The following Hamiltonian, $H_{1/2}(\bm{k})$ in Eq. \ref{8:main} are the symmetry-allowed $\bm{k \cdot p}$ models around $\Gamma$ point for $M$Se$_2$, $M$S$_2$ and $M$Te$_2$, respectively (Complete symmetry-allowed terms and detail explanations in Appendix.D).

\begin{equation} \label{8:main}
\begin{aligned}
  H_1(\bm{k}) &= \varepsilon_1(\bm{k})I_4 + M_1(\bm{k})s_z\sigma_0 +  \nu_1 k_x s_x \sigma_y    \\
              &  + \nu_2 k_x s_x \sigma_z, \\
  H_2(\bm{k}) &= \varepsilon_2(\bm{k})I_6 + M_2(\bm{k})\lambda_3\sigma_0 + M_3(\bm{k})\lambda_8 \sigma_0 \\
             & + \nu_1 k_x \lambda_7 \sigma_0 + \nu_2 k_x \lambda_4 \sigma_z, \\
\end{aligned}
\end{equation}
where
\begin{equation}
\begin{aligned}
  \varepsilon_{1/2}(\bm{k}) &= \mu  + t_{x,1} k_x^2 + t_{y,1} k_y^2, \nonumber \\
  M_{1/2}(\bm{k}) &= \delta + t_{x,2} k_x^2 + t_{y,2} k_y^2, \nonumber  \\
  M_3(\bm{k}) &= \delta' + t_{x,3} k_x^2 + t_{y,3} k_y^2 \nonumber .
\end{aligned}
\end{equation}

\begin{table}[htbp]
\begin{tabular}{lllllll}
\hline\hline
  &WTe$_2$ &MoTe$_2$  &WSe$_2$ &MoSe$_2$ &WS$_2$ &MoS$_2$\\
\hline
$\delta$ (eV)   & \ 0.470 &\ 0.180 &\ 0.342 &\ 0.363 &\ 0.094 &\ 0.239  \\
$\delta'$(eV)   & \ 0.495 &\ 0.443 & \ \ \ \ -- & \ \ \ \ -- & \ \ \ \ -- &  \ \ \ \ -- \\
$\mu$ (eV)      &\ 0.205 &\ 0.086 &\ 0.319 &\ 0.325 &\ 0.051 &\ 0.136  \\
$t_{x,1}$ (eV $\cdot$ \AA $^2$ ) & -7.363 & -7.535  & -14.814 & -12.723 & -3.202 & -6.826 \\
$t_{y,1}$ (eV $\cdot$ \AA $^2$ ) & -4.928 & -4.656  & -3.777 & -4.704 & -3.732 & -2.383 \\
$t_{x,2}$ (eV $\cdot$ \AA $^2$ ) & -7.207 & -5.756  & -16.107 & -14.148 & -5.869 & -9.644 \\
$t_{y,2}$ (eV $\cdot$ \AA $^2$ ) & -0.184 &\ 0.519  &\ 1.137 & -0.149 & \ 7.269 &\ 3.414 \\
$t_{x,3}$ (eV $\cdot$ \AA $^2$ ) & -7.996 & -8.665  &  \ \ \ \ -- & \ \ \ \ -- & \ \ \ \ -- &  \ \ \ \ -- \\
$t_{y,3}$ (eV $\cdot$ \AA $^2$ ) & -3.223 & -3.282 &  \ \ \ \ -- & \ \ \ \ -- & \ \ \ \ -- &  \ \ \ \ -- \\
$\nu_1$ (eV $\cdot$ \AA  )   &\ 0.699 &\ 1.157  &\ 0.168 & \ 0.343 & \  0.387 & \ 0.337 \\
$\nu_2$ (eV $\cdot$ \AA  )  &\ 0.349 &\ 0.358 &\ 0.333 &\ 0.343 &\ 0.482 &\ 0.153  \\
\hline\hline
\end{tabular}
\caption{ Fitted $\bm{k\cdot p}$ model parameters for all six $MX_2$.}
\label{TABLE 4}
\end{table}

In Eq. \ref{8:main}, we only keep $k_i^2$($i=x,y$) in the mass dependent terms and $k_i$ in the velocity terms. The positive $\delta$ and $\delta'$ represent the band inversion nature at $\Gamma$ point. Without SOC, the band crossing only occurs along $X$ - $\Gamma$, thus we ignore $k_y$ terms in the $\bm{k \cdot p}$ model. In this sense, our $\bm{k \cdot p}$ model at least can capture the band shape and the orbital reordering nature.

The $\bm{k \cdot p}$ model for $M$Te$_2$ reveals the interplays between three orbitals and provides another angle to understand the semimetallic properties of $M$Te$_2$. In TABLE \ref{TABLE 4}, the $t_x$ terms are much larger than the $t_y$ terms, suggesting a strong anisotropy in momentum space. It is also found that by fine tuning the SOC strength, a positive gap opens and the gap value shows a monotonic relationship with the SOC strength. In the next section, we discuss a strained TB model, where the strain effects are reflected in the hopping parameters.

\section{STRAINED TB MODEL} \label{strain}

Strain is one of the most important approaches to tune the electronic properties of 2D materials. Experimental results prove the strain can induce metal-insulator transition and enhance the metastable phase growth\cite{chen2021epitaxial,zhao2020strain}. In this section, we introduce the strained TB model based on the $\uppercase\expandafter{\romannumeral1}$-TB model.

\begin{figure}
\includegraphics[width=\columnwidth]{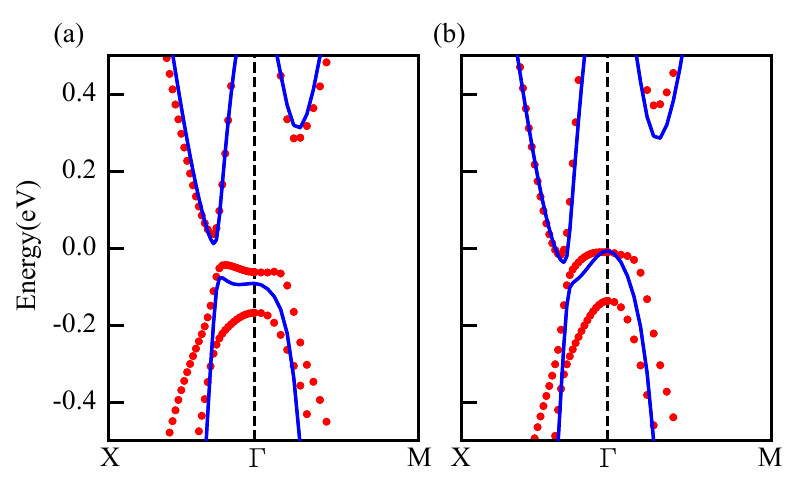}
\caption{\label{Fig.strain} The Band structures from first-principles calculations and strained TB model. (a) and (b) are WTe$_2$ band structures under {a} 5\% tensile strain along the $\bm{R_x}$ direction and under a compressive strain along the $\bm{R_y}$ direction, respectively. All the first-principles calculation results and strained TB model results are labeled by red dots and blue lines, respectively.}
\end{figure}
Assuming that a small uniaxial strain (around $\pm$ 1.0\%) is applied to $MX_2$ without breaking any symmetries, we consider the strain as a small perturbation and only change the overlap between different orbitals. In the strained TB model, the strain effects on band structures are captured by changing the hopping parameters. Then a series of linear fittings are conducted within a small range of strain using Eq. \ref{equ:strain}: \\
\begin{equation} \label{equ:strain}
\begin{aligned}
  t'(r_{ix}(1+\Delta_x|\bm{R_x}|),r_{iy}) &=  \sum_{j=0} a_{x,j}\Delta_x^j, \\
  t'(r_{ix},r_{iy}(1+\Delta_y|\bm{R_y}|)) &=  \sum_{j=0} a_{y,j}\Delta_y^j.
\end{aligned}
\end{equation}

Here, only a uniaxial strain is included. $a_{x/y,j}$ represents the $j^{th}$ order polynomial coefficient in the $x/y$ direction and $\Delta_{x/y}$ is the strain strength.

\begin{table*}
\begin{tabular}{p{1.5cm} p{1.5cm} p{1.5cm} p{1.5cm} p{1.5cm} p{1.5cm} p{1.5cm} p{1.5cm}}
\hline\hline
 $a_{x,1}(t_1)$ &$a_{x,0}(t_1)$ &$a_{x,1}(d_1)$ &$a_{x,0}(d_1)$ & $a_{x,1}(p_1)$ & $a_{x,0}(p_1)$ & $a_{x,1}(p_2)$ &$a_{x,0}(p_2)$ \\
 $a_{y,1}(t_1)$ &$a_{y,0}(t_1)$ &$a_{y,1}(d_1)$ &$a_{y,0}(d_1)$ & $a_{y,1}(p_1)$ & $a_{y,0}(p_1)$ & $a_{y,1}(p_2)$ &$a_{y,0}(p_2)$ \\
  \hline
          \  -3.059 &\ 1.019 & -1.467 &\ 0.468 & -2.207 &\ 1.063 & -2.798 &\ 0.374  \\
          \ \ 4.823 &\ 1.015 &\ 0.330 &\ 0.461 &\ 2.777 &\ 1.068 &\ 4.225 &\ 0.376  \\
\hline\hline
\end{tabular}
\caption{Fitted parameters for WTe$_2$ strained TB model.}
\label{TABLE 5}
\end{table*}

In Fig. \ref{Fig.strain}, we present the band structures of 1T'-WTe$_2$ under uniaxial strain along the $\bm{R_x}$ and the $\bm{R_y}$ directions, respectively. Our strained TB model results for 1T'-WTe$_2$ are based on Eq. \ref{equ:strain} to the linear order which are consistent with the first-principles calculations up to ~5\% strain strength. The strain effects along different direction are not identical. To be more specific, it is the tensile strain along the $x$ direction or the compressive strain along the $y$ direction that contributes to a more positive fundamental gap. Our strained TB model also reflects this strain anisotropy: in TABLE \ref{TABLE 5}, most linear order parameters have different signs along different directions. Besides, we believe the bulk conductance of all six $MX_2$ materials can be easily manipulated by external strain due to the similar orbital components and dispersions. Our strained TB model serves as a suitable starting point to analyze the electronic properties of $MX_2$ under perturbations.

In Appendix E, we provide the insulating 1T'-WTe$_2$ nontrivial edge density of states under strong enough uniaxial strain. This shows that the strain does not only open a fundamental gap in $MX_2$, but also preserves the nontrivial topology.

\section{SUMMARY} \label{summary}

In this paper, we constructed four-band TB models based on the orbital components of monolayer 1T'-$MX_2$ . By analyzing the nonsymmorphic space group of 1T'-$MX_2$ structure, bases from $A_g$ and $A_u$ IRs in $C_{2h}$ were selected. Our TB models match well with the first-principles band structures along all high symmetric lines.
$\uppercase\expandafter{\romannumeral1}$-TB model with fewer parameters fits well within the energy window from -0.3 eV to 0.3 eV; $\uppercase\expandafter{\romannumeral2}$-TB model with more parameters fits well within the energy window from -0.3 eV to 0.8 eV.
With the inclusion of SOC effects, both TB models can capture the nontrivial topological properties of 1T'-$MX_2$.
Specifically, we find that the momenta belong to the same little group have different degeneracies due to features of the nonsymmorphic group. Plus, the spin of edge states along the $\bm{R_x}$ direction always lies in the $y-z$ plane, which is guaranteed by the time-reversal symmetry and the mirror symmetry $M_x$.
Moreover,  for the first time, we found that $MX_2$ have different parities at $\Gamma$ point, and then we built two different types of $\bm{k\cdot p}$ model to fully describe the low-energy properties of $MX_2$.
In the last section, we built the strained TB model to capture the effects of uniaxial strain along $\bm{R_x}$ and $\bm{R_y}$. The small tensile strain along the $x$ direction or compressive strain along the $y$ direction induces a metal-insulator transition. To conclude, our TB and $\bm{k\cdot p}$ models are simple and efficient to describe the 1T'-$MX_2$ band structure and its topological properties, which can serve as a solid foundation for future studies of superconductivity and other many-body physics in 1T'-$MX_2$.

\section{ACKNOWLEDGMENTS}
We thank Vic K. T. Law, Jason Z.S. Gao, and Noah F.Q. Yuan for helpful discussions. This work is supported by the Research Grants Council of Hong Kong (ECS26302118,16305019, 16306220 and N\_HKUST626/18) and National Natural Science Foundation of China (NSFC20SC07).

\appendix
\setcounter{secnumdepth}{0}

\section{Appendix A: Layout of Tight-binding parameters}

\begin{figure}[htbp]
\includegraphics{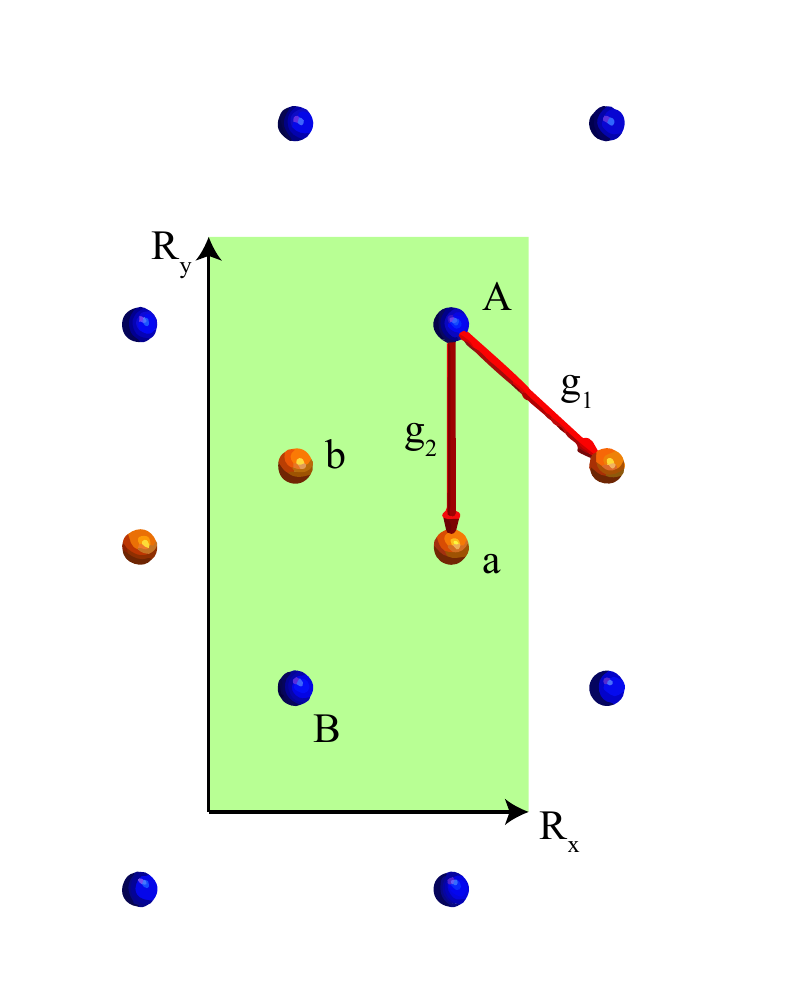}
\caption{Top view of the reduced lattice structure of monolayer $MX_2$. $M$ atom is in blue and $X_1$ atom is in orange. Unit cell is labeled by the green rectangle. Two basic vectors are labeled as $\bm{g_1}$and $\bm{g_2}$ to be accounted for generating other $\bm{r_{i}}$. The unit vectors are labeled as $\bm{R_x}$ and $\bm{R_y}$}
\label{fig:nnd}
\end{figure}
We label all the hoppings in real space included in our TB model with the following notations.

\begin{equation}
\nonumber
\begin{aligned}
&\bm{r^{dp}_1}=\bm{g_1} , & \bm{r^{dp}_2}&=\bm{g_2} + \bm{R_x} \\
&\bm{r^{dp}_3}=\bm{g_1} + \bm{R_x} + \bm{R_y}, & \bm{r^{dp}_4}&=\bm{g_2} + \bm{R_x} + \bm{R_y} \\
&\bm{r^{dp}_5}=\bm{g_2} + \bm{R_x} - \bm{R_y} , &  \bm{r^{dp}_6}&=\bm{g_1} + \bm{R_x} - \bm{R_y}\\
&\bm{r^{pp}_{1}}= \bm{R_x} , &\bm{r^{pp}_{2}} &= \bm{g_1} - \bm{g_2} \\
&\bm{r^{pp}_{3}}= \bm{R_y} ,  &\bm{r^{pp}_{4}} &= \bm{g_1} - \bm{g_2} -\bm{R_y} \\
&\bm{r^{pp}_{5}}= \bm{g_1} - \bm{g_2} + \bm{R_y} , &\bm{r^{pp}_{6}}&= \bm{R_x} + \bm{R_y}\\
&\bm{r^{dd}_{1}}= \bm{g_1} + \bm{g_2} + \bm{R_y} , &\bm{r^{dd}_{2}}&= \bm{R_x} \\
&\bm{r^{dd}_{3}}= \bm{g_1} + \bm{g_2} , &\bm{r^{dd}_{4}}&= \bm{R_y} \\
&\bm{r^{dd}_{5}}= \bm{R_x} + \bm{R_y} , &\bm{r^{dd}_{6}}&= \bm{g_1} + \bm{g_2} - \bm{R_y}\\
\end{aligned}
\end{equation}

The hopping parameters in the main text are defined as
\begin{equation}
\begin{aligned}
t_i( \bm{r^{dp}_i} )&= \mel{\phi_{l}(\bm{r})} {\hat H} {\phi_{l'}(\bm{r - r^{dp}_i}) } ; \nonumber \\ d_i(\bm{r^{dd}_i})&= \mel{\phi_{l}(\bm{r})}{\hat H}{\phi_{l'}(\bm{r-r^{dd}_i})} ; \nonumber \\ p_i(\bm{r^{pp}_i})&= \mel{\phi_{l}(\bm{r})}{\hat H}{\phi_{l'}(\bm{r - r^{pp}_i})} (l,l' = A,B,a,b).  \nonumber \\
\end{aligned}
\end{equation}

The symmetry constraints of the hopping parameters read:
\begin{equation}
\begin{aligned}
&t_i(\bm{r^{dp}_i}) =&- t_i(M_x\bm{r^{dp}_i}) = & - t_i(P\bm{r^{dp}_i}) \\
&d_i(\bm{r^{dd}_i}) =&\ d_i(M_x\bm{r^{dp}_i}) =& \ d_i(P\bm{r^{dd}_i}) \\
&p_i(\bm{r^{dd}_i}) =&\ p_i(M_x\bm{r^{dp}_i}) =& \ p_i(P\bm{r^{dd}_i}) \\
\end{aligned}
\end{equation}

Similarly, we consider the symmetry constraints in the SOC terms. In the main text, only $\bm{g_1},\ \bm{g_2}$ connected off-site SOC terms are constructed. We label the SOC parameters as: $il_j( \bm{g_j} )=\mel{\phi_{A/B,\alpha}(\bm{r})} {\hat H} {\phi_{a/b,\alpha '}(\bm{r - g_j}) } $,\ ($\alpha,\alpha ' = \uparrow,\downarrow$).

Since the $M_x$ operation anti-commutes with $\sigma_y$ and $\sigma_z$ in spin bases, we consider the SOC effects in the $y$ and the $z$ directions. The symmetry constraints to the SOC parameters read:
\begin{equation}
\begin{aligned}
 &il_j( \bm{g_1} )= \ il_j( M_x\bm{g_1} ) &= -il_j( P\bm{g_1} ) \\
 &il_j( \bm{g_2} )= -il_j( P\bm{g_2} ) & \\
\end{aligned}
\end{equation}

\subsection{Appendix B: Edge density of states}
\begin{figure}[htbp]
\includegraphics{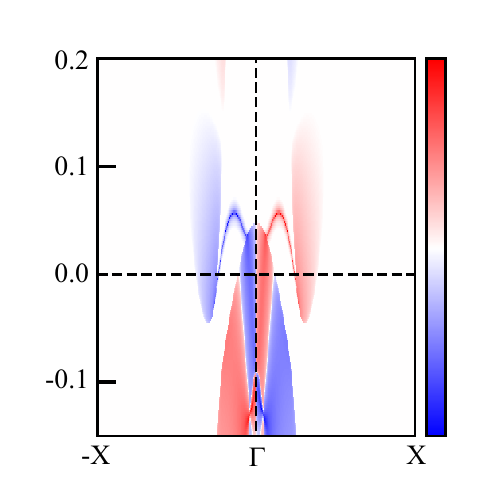}
\caption{WTe$_2$ edge spin-polarized density of states in the $y$ direction.}
\label{fig:spiny}
\end{figure}

In the main-text, spin-polarized density of states in the $z$ direction is presented. Here we supplement the result in the $y$ direction.

\subsection{Appendix C: Symmetry-protected degeneracy and zero spin polarization}

For any lattice structure which preserves time-reversal ($T$) and inversion ($P$) symmetries, the Hamiltonian satisfies the constraints:

\begin{equation}
\begin{aligned}
[ H,T] &= [H , P ] = 0      \\  \label{eq:1}
H \ket{k_i,\sigma_i}  &= E \ket{k_i,\sigma_i}
\end{aligned}
\end{equation}

where $\ket{k_i,\sigma_i}$ is the eigenstate of $H$, we apply the symmetry operations ($T,P$) on both sides of the equation:
\begin{equation}
\begin{aligned}
  TH \ket{k_i,\sigma_i} &= ET \ket{k_i,\sigma_i}  \\
  PH \ket{k_i,\sigma_i} &= EP \ket{k_i,\sigma_i}   \\
\text{with} \eqref{eq:1},  \\
  TH \ket{k_i,\sigma_i} &= HT \ket{k_i,\sigma_i}  \\
  PH \ket{k_i,\sigma_i} &= HP \ket{k_i,\sigma_i}
\end{aligned}
\end{equation}

now for every eigenvalue $E$, we have at most two eigenstates labeled by different spin vectors:
\begin{equation}
\begin{aligned}
  T \ket{k_i,\sigma_i} &= \ket{-k_i,-\sigma_i} \\
  P \ket{k_i,\sigma_i} &= \ket{-k_i, \sigma_i}
\end{aligned}
\end{equation}

Here, we denote the eigenstates by momentum and spin polarization ($\ket{k_i,\sigma_i},\ (i = x, y, z)$). For any spin polarization direction, if $\sigma_i$ is not zero, we can always find the spin symmetry `partner' ($-\sigma_i$) and the summation of every spin partner ($\rho(E,k_i)_i$) is zero; if $\sigma_i$ is zero, it will directly contribute to the zero spin density of states. With a similar deduction process, when the edge structure only preserves mirror ($M$) and time-reversal symmetries, since mirror symmetry operator ($M_i = i\sigma_i$) commute with $\sigma_i$, the spin symmetry `partner' ($-\sigma_i$) can only be found along the $i$ direction. To summarize, the zero $\rho(E,k_i)_i$ protected by the combination of time-reversal symmetry and one crystalline symmetry whose operator commute with $\sigma_i$.

\subsection{Appendix D: symmetry constraints of the \lowercase{\textit{k}}$\cdot$\lowercase{\textit{p}} model}
We use the bases \{$\ket{\phi_{p_x},\uparrow},\ket{\phi_{p_x},\downarrow},\ket{\phi_{d_{xz},\uparrow}},$$\ket{\phi_{d_{xz}},\downarrow}$\} for $M$Se$_2$, $M$S$_2$ and \{$\ket{\phi_{d_{z^2}},\uparrow},\ket{\phi_{d_{z^2}},\downarrow},\ket{\phi_{d_{xz}},\uparrow},\ket{\phi_{d_{xz}},\downarrow},$ $\ket{\phi_{p_{y},\uparrow}},\ket{\phi_{p_{y}},\downarrow}$\} for $M$Te$_2$.

Symmetry operations at $\Gamma$ point are $M_x = i s_z \sigma_x$, $P = s_z  \sigma_0$, and $T = i s_0  \sigma_y K$ for $M$Se$_2$ and $M$S$_2$, and $P = (\lambda_0 + 2\sqrt{3} \lambda_8) \sigma_0$, $M_x = i(\lambda_0 + 3\lambda_3 - \sqrt{3} \lambda_8) \sigma_x$ and $T = i \lambda_0 \sigma_y K$ for $M$Te$_2$. The following table lists the symmetry transformation relationships under different $k_i$, $s_i \sigma_j$, and $\lambda_i \sigma_j$.\\

\begin{table}[htbp]
\begin{tabular}{|l|l|l|l|l|l|l|}
\hline
  $P$ &$M_x$  & $T$   & $\bm{k}$ &$s_i \sigma_i$ & $\lambda_i \sigma_i$ \\
\hline
   +1 & +1 & +1 & 1;$k_x^2$;$k_y^2$  & $s_{0/z}\sigma_0$  & $\lambda_{3/8} \sigma_0$ ;$\lambda_{2} \sigma_{y/z}$ \\
   +1 & +1 & -1 &  \ \ \ \ --       & $s_{0/z}\sigma_x$  & $\lambda_{3/8} \sigma_x$ ;$\lambda_{1} \sigma_{y/z}$  \\
   +1 & -1 & +1 & $k_xk_y$           &  \ \ \ \ --                      & $\lambda_{1} \sigma_0$;$\lambda_{2} \sigma_x$ \\
   -1 & +1 & +1 &  \ \ \ \ --       & $s_{x}\sigma_0$ ; $s_{y}\sigma_x$ & $\lambda_{4} \sigma_0$;$\lambda_{5} \sigma_x$;$\lambda_{7} \sigma_{y/z}$\\
   +1 & -1 & -1 &  \ \ \ \ --       & $s_{0/z}\sigma_{y/z}$             & $\lambda_{2} \sigma_0$;$\lambda_{1} \sigma_x$;$\lambda_{3/8} \sigma_{y/z}$ \\
   -1 & +1 & -1 & $k_y$              & $s_{x}\sigma_x$ ; $s_{y}\sigma_0$ & $\lambda_{5} \sigma_0$;$\lambda_{4} \sigma_x$;$\lambda_{6} \sigma_{y/z}$ \\
   -1 & -1 & +1 &  \ \ \ \ --       & $s_{y}\sigma_{y/z}$               & $\lambda_{6} \sigma_0$;$\lambda_{7} \sigma_x$;$\lambda_{5} \sigma_{y/z}$ \\
   -1 & -1 & -1 & $k_x$              & $s_{x}\sigma_{y/z}$               & $\lambda_{7} \sigma_0$;$\lambda_{6} \sigma_x$;$\lambda_{4} \sigma_{y/z}$ \\
\hline
\end{tabular}
\caption{The character table for the $\bm{k \cdot p}$ models.}
\label{TABLE kp}
\end{table}

The symmetry-allowed terms in the $\bm{k \cdot p}$ model read:
\begin{equation} \label{kp}
\begin{aligned}
 H_1(\bm{k}) & = \epsilon_1(\bm{k}) I_4  + M_1(\bm{k}) s_{z}\sigma_{0} + \nu_1 k_x s_x \sigma_y + \nu_2 k_x s_x \sigma_z\\
            & + \nu_3 k_y s_x \sigma_x + \nu_4 k_y s_y \sigma_0 \\
 H_2(\bm{k}) & = \epsilon_2(\bm{k}) I_6  + M_1(\bm{k}) \lambda_{3}\sigma_{0} + M_2(\bm{k}) \lambda_{8}\sigma_{0} \\
             & + M_3(\bm{k}) \lambda_{2}\sigma_{y}   + M_4(\bm{k}) \lambda_{2}\sigma_{z}  + \nu_1 k_x \lambda_{7}\sigma_{0} \\
             & + \nu_2 k_x \lambda_{4}\sigma_{z} + \nu_3 k_x \lambda_{4}\sigma_{y}  + \nu_4 k_x \lambda_{6}\sigma_{x} \\
             &+ \nu_5 k_y \lambda_{5}\sigma_{0} + \nu_6 k_y \lambda_{4}\sigma_{x} + \nu_7 k_y \lambda_{6}\sigma_{y}\\
            & + \nu_8 k_y \lambda_{6}\sigma_{z} \\
 \epsilon_{1/2}(\bm{k}) &= \mu + t_{x,1} k_x^2 + t_{y,1} k_y^2 \\
 M_1(\bm{k})  &= -\delta + t_{x,2} k_x^2 +t_{y,2} k_y^2 \\
 M_2(\bm{k})  &= -\delta' + t_{x,3} k_x^2 +t_{y,3} k_y^2 \\
 M_3(\bm{k})  &= -\delta_3 + t_{x,4} k_x^2 +t_{y,4} k_y^2 \\
 M_4(\bm{k})  &= -\delta_4 + t_{x,5} k_x^2 +t_{y,5} k_y^2 \\
\end{aligned}
\end{equation}
In Eq. \ref{kp} we list all the symmetry-allowed terms up to $k_i^2$ ($i = x,y$). In the main text, only $\epsilon_{1/2}$, $M(\bm{k})_{1/2}$, and $\nu_{1/2}$ are included.

\subsection{Appendix E: Edge density of states from strained TB model}

\begin{figure}[htbp]
\includegraphics{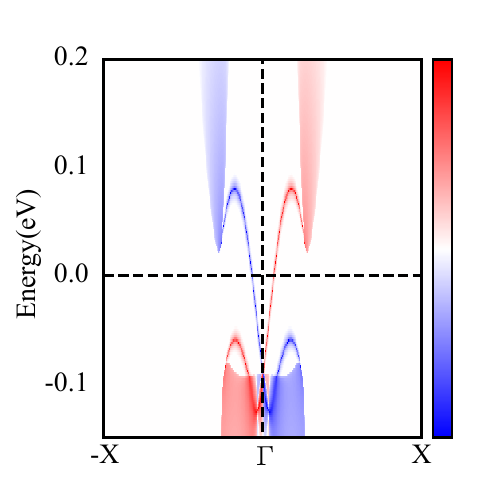}
\caption{WTe$_2$ edge spin-polarized density of states in the $y$ direction. The strained TB parameters are under 5\% tensile strain along $\bm{R_x}$ direction.}
\label{fig.spiny}
\end{figure}

\clearpage


%

\end{document}